\documentclass[12pt]{article}

\usepackage{graphicx}
\usepackage{a4}

\usepackage{fancybox}
\usepackage{epsfig}
\usepackage{color}

\usepackage{amsfonts}
\usepackage{mathrsfs}
\usepackage{amssymb}
\usepackage{amsmath}

\usepackage{dcolumn}
\usepackage{bm}

\usepackage{float}

\def\pa{\partial}

\def\be{\beta}
\def\ga{\gamma}

\def\th{\theta}

\def\ka{\kappa}

\def\Si{\Sigma}

\def\Om{\Omega}

\newcommand{\ben}{\begin{equation}}
\newcommand{\een}{\end{equation}}
\newcommand{\bea}{\begin{eqnarray}}
\newcommand{\eea}{\end{eqnarray}}
\newcommand{\ba}{\begin{array}}
\newcommand{\ea}{\end{array}}
\newcommand{\bit}{\begin{itemize}}
\newcommand{\eit}{\end{itemize}}

\newcommand{\vs}[1]{\vspace{#1 mm}}

\newcommand{\dsl}{\pa \kern-0.5em /}

\begin{document}

\topmargin 0pt \oddsidemargin 0mm




\vspace{2mm}

\begin{center}

{\Large \bf Accelerating solitons in Rindler space}

\vs{10}

 {\large Huiquan Li \footnote{E-mail: hli@qzc.edu.cn} and Jianyong Wang \footnote{E-mail: 32040@qzc.edu.cn}}

\vspace{6mm}

{\em

College of Teacher Education, Quzhou University, \\
Quzhou, Zhejiang 324000, China}

\end{center}

\vs{9}

\begin{abstract}

Relativistic nonlinear field theories that are integrable in two-dimensional Minkowski spacetime remain integrable in Rindler space. We investigate the properties of classical sine-Gordon (SG) solitons in Rindler space. Accelerating observers in each of the two Rindler wedges can see only a portion of the full soliton in Minkowski spacetime; this portion itself forms a soliton carrying a partial topological charge. The soliton solution admits the following interpretation: a soliton incoming from the horizon decelerates to stop at some point, then falls back and accelerates to the speed of light toward the horizon. Energy can flow into and out of the Rindler horizon. The higher energy soliton gains energy from the horizon while the lower one loses energy.

\end{abstract}




\section{Introduction}
\label{sec:}

Solitons are particle-like wave packets that arise as exact solutions of a wide class of nonlinear field equations, which are typically obtained in Minkowski spacetime \cite{Manton:2004tk,Shnir:2018yzp}. Such solitons usually propagate at a constant velocity. A notable example is provided by the sine-Gordon (SG) solitons, which are widely studied in various contexts.

When solitons propagate through inhomogeneous media, their effective mass may become spacetime dependent. Consequently, the soliton velocity is no longer constant but varies with position and time. The inhomogeneity of the medium or the presence of external forces usually destroys the integrability of a system, making analytic study difficult \cite{Kivshar:1989ue}. Nevertheless, there exists a class of variable-mass sine-Gordon (VMSG) models \cite{Kundu:2007jh, Hoare:2020fye, Aguirre:2026jbx} that remain integrable at both the classical and quantum levels, admitting exact soliton solutions that can accelerate and change their shape, width, and amplitude. Accelerating solitons in other models and contexts have also been discussed in \cite{Gorbach_2007, Kundu_2008, Fac_o_2013, Mukherjee_2019, Qin_2019, Sakaguchi_2019, Melnikov:2020ret, Malomed_2022}. 

In this work, we study SG solitons in Rindler space, which captures the near-horizon geometry of non-extreme black holes. Since the SG theory is relativistic, it can be consistently formulated in a curved spacetime. The SG equation in Rindler space takes the same form as that in an inhomogeneous medium, and it remains exactly solvable. This therefore provides a useful setting for understanding the behaviour of a structured object instead of a point particle near a black hole. The Rindler wedge, being only a portion of the full Minkowski spacetime, introduces a natural notion of partial observation: an accelerating observer cannot access the entire soliton configuration, and the portion visible within the wedge itself constitutes a soliton. This expectation raises interesting questions about how the solitons propagate in the accelerating frame, how topological information is partitioned across the horizon and how energy is distributed between the two Rindler wedges.

Analyses of quantum field theory in Rindler space lead to the Unruh effect, which is essential for understanding Hawking radiation from black holes (e.g., see \cite{Birrell:1982ix, Fulling:1989nb}). 
The analyses are usually carried out within linear field equations, where the mode decomposition and the Bogoliubov transformation between Minkowski and Rindler modes can be performed explicitly. Here, the integrable SG model in Rindler space offers a rare opportunity to examine horizon physics beyond the linear regime. 

The paper is organised as follows. We first sumarrize the results of solitons in two-dimensional Minkowski spacetime in Section.\ (\ref{sec:Minsol}). Then we present the soliton solutions in Rindler space in Section.\ (\ref{sec:Rinsol}) and discuss their properties in Section.\ (\ref{sec:topcharge}), (\ref{sec:energy}) and (\ref{sec:tempasy}). Finally, we give some discussions on the results in the last section.

\section{SG solitons in Minkowksi spacetime}
\label{sec:Minsol}

In the SG theory, we consider a scalar field with the potential
\begin{equation}\label{e:SGpot}
    V(\phi)=m_0^2(1-\cos\phi).
\end{equation}
where the mass $m_0$ is a constant. The resulting SG equation reads:
\begin{equation}
    -\phi_{tt}+\phi_{xx}-m_0^2\sin\phi=0.
\end{equation}

The quation is solvable and has two kinds of soliton solutions, the kink one and the anti-kink one
\begin{equation}
    \phi_{\text{K}}(t, x)=4\arctan e^{\ga-\ga_0}, \quad \phi_{\bar{\text{K}}}(t, x)=4\arctan e^{-\ga+\ga_0},
\end{equation}
where 
\[\ga_0=m_0\cosh\th_0x_0,\]
\[\ga=m_0(\cosh\th_0x-\sinh\th_0 t)\in(-\infty,\infty).\]
The kink solution varies from $\phi_{\text{K}}(x\rightarrow-\infty)=0$ to $\phi_{\text{K}}(x\rightarrow\infty)=2\pi$ along the $x$-direction. So it has topological charge $Q=1$. The anti-kink has charge $Q=-1$, with the field from $\phi_{\bar{\text{K}}}(x\rightarrow-\infty)=2\pi$ to $\phi_{\bar{\text{K}}}(x\rightarrow\infty)=0$. The propagation speed of the soliton is constant
\[v_0=\tanh\th_0.\] 

The compoenents of the energy-momentum tensor for the kink solution are
\begin{equation}
    T_{tt}=4m_0^2\cosh^2\th_0\operatorname{sech}^2(\ga-\ga_0),
\end{equation}
\begin{equation}
    T_{tx}=T_{xt}=-2m_0^2\sinh(2\th_0)\operatorname{sech}^2(\ga-\ga_0),
\end{equation}
\begin{equation}
    T_{xx}=4m_0^2\sinh^2\th_0\operatorname{sech}^2(\ga-\ga_0).
\end{equation}
The components are all of a lump shape. On a fixed time slice, the total energy and momentum of the system can be obtained from the integrals:
\begin{equation}
    E=\int T_{tt}dx=\frac{8m_0}{\sqrt{1-v_0^2}}, \quad P=\int T_{tx}dx=-\frac{8m_0v_0}{\sqrt{1-v_0^2}}.
\end{equation}
They describe a particle-like object with constant mass and momentum. Obviously, they diverge at the speed of light. It is easy to find that the total fluxes of  charge and energy through any point $x$ are
\begin{equation}
    F_c=-\frac{1}{2\pi}\int \pa_t\phi dt=Q, \quad  F_e=-\int T_{tx}dt=E.
\end{equation}
The relations indicate that the whole energy and charge of the soliton will pass through every point and will propagate to infinity eventually.

The SG model with variable mass emergences in various physical systems \cite{Kundu:2007jh, Hoare:2020fye, Aguirre:2026jbx}.
In \cite{Kundu:2007jh}, it is found that the equation of the following form 
\begin{equation}\label{e:varSGeq}
    \phi_{uv}-m_0^2f(u)g(v)\sin\phi=0,
\end{equation}
is exactly integrable. Here $(u,v)$ are the lightcone coordinates
\begin{equation}
    u=\frac{1}{2}(x+t), \quad v=\frac{1}{2}(x-t).
\end{equation}
Since the mass varies with spacetime coordinates, the shape and velocity of the soliton are now possbily changing.

\section{Solitons in Rindler space}
\label{sec:Rinsol}

The SG equation in Rindler space takes the form (\ref{e:varSGeq}). The Rindler space is the universal geometry near non-extreme black holes. For example, the near-horizon geometry ($|\rho|\ll1$) of a Schwarzschild or non-extreme RN black hole can be expressed as Rindler space times sphere $S^2$
\[ds^2=-\rho^2d\tau^2+\frac{1}{\be^2}d\rho^2+r_+^2d\Om^2,\] 
where the constant $\be$ is related to the radius $r_+$ of the event horizon. The metric of the Rindler space can be alternatively expressed as
\begin{equation}
    ds_R^2=e^{-2\be R}(-d\tau^2+dR^2)=4e^{-2\be(\xi^++\xi^-)}d\xi^+d\xi^-.
\end{equation}
where the light-cone cordinates are
\begin{equation}
    \xi^+=\frac{1}{2}(R+\tau), \quad \xi^-=\frac{1}{2}(R-\tau).
\end{equation}  
They are related to the ones in Minkoski spacetime via
\begin{equation}
    u=\pm\frac{1}{2\be}e^{-2\be\xi^-}, \quad v=\pm\frac{1}{2\be}e^{-2\be\xi^+},
\end{equation}
The $\pm$ signs represent the right and left Rindler wedges, respectively. The horizon is located at $R\rightarrow\infty$ for either case. Thus, the Rindler wedges only conver a portion of the Minkowksi spacetime. Observers move with constant proper acceleration in Rindler space.

On the near-horizon geometry, the field equation for a scalar field $\phi$ with potential $V(\phi)$ is
\[-\phi_{\tau\tau}+\phi_{RR}-e^{-2\be R}\left[\frac{1}{r_+^2}l(l+1)\phi+V'(\phi)\right]=0.\]
for the $l$-th order harmonic mode on $S^2$. In this work, we consider the $l=0$ case with the SG potential (\ref{e:SGpot})
\begin{equation}
    \pa_+\pa_-\phi-m_0^2e^{-2\be(\xi^++\xi^-)}\sin\phi=0.
\end{equation}
We discuss the SG equation in the whole Rindler space, but not restricted to the near horizon ($|\rho|\ll1$) region.

The solution of the equation can be found immediately with a coordinate transformtion. But the redefintion of the coordinates lead to the change of the domain and boundary conditions of the field. Since each Rindler wedge only covers a portion of the full Minkowksi spacetime, the kink solutions in the two wedges are
\begin{equation}
    \phi^{(R)}_{\text{K}}(\tau, R)=4\arctan e^{\ga-\ga_0},
\end{equation}
\begin{equation}
    \phi^{(L)}_{\text{K}}(\tau, R)=4\arctan e^{-\ga-\ga_0},
\end{equation}
where $\ga_0=m_0\cosh\th_0x_0$ is a constant and 
\begin{equation}
    \ga=\frac{m_0}{\be}e^{-\be R}\cosh(\be\tau-\th_0)\in(0,\infty).
\end{equation}
So the variable $\ga$ here only takes positive values.

\begin{figure}[H]
    \center
    \includegraphics[width=0.48\columnwidth]{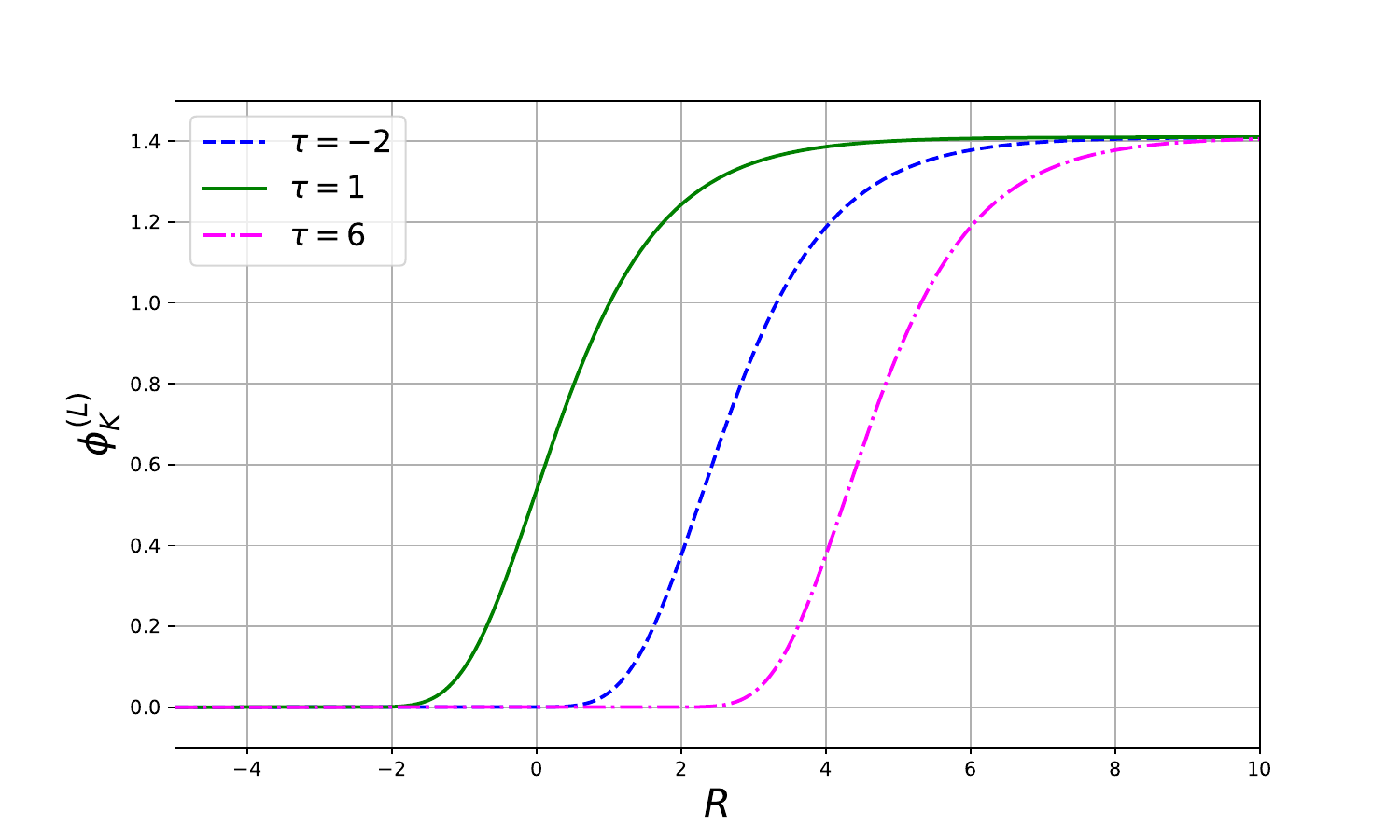}
    \includegraphics[width=0.48\columnwidth]{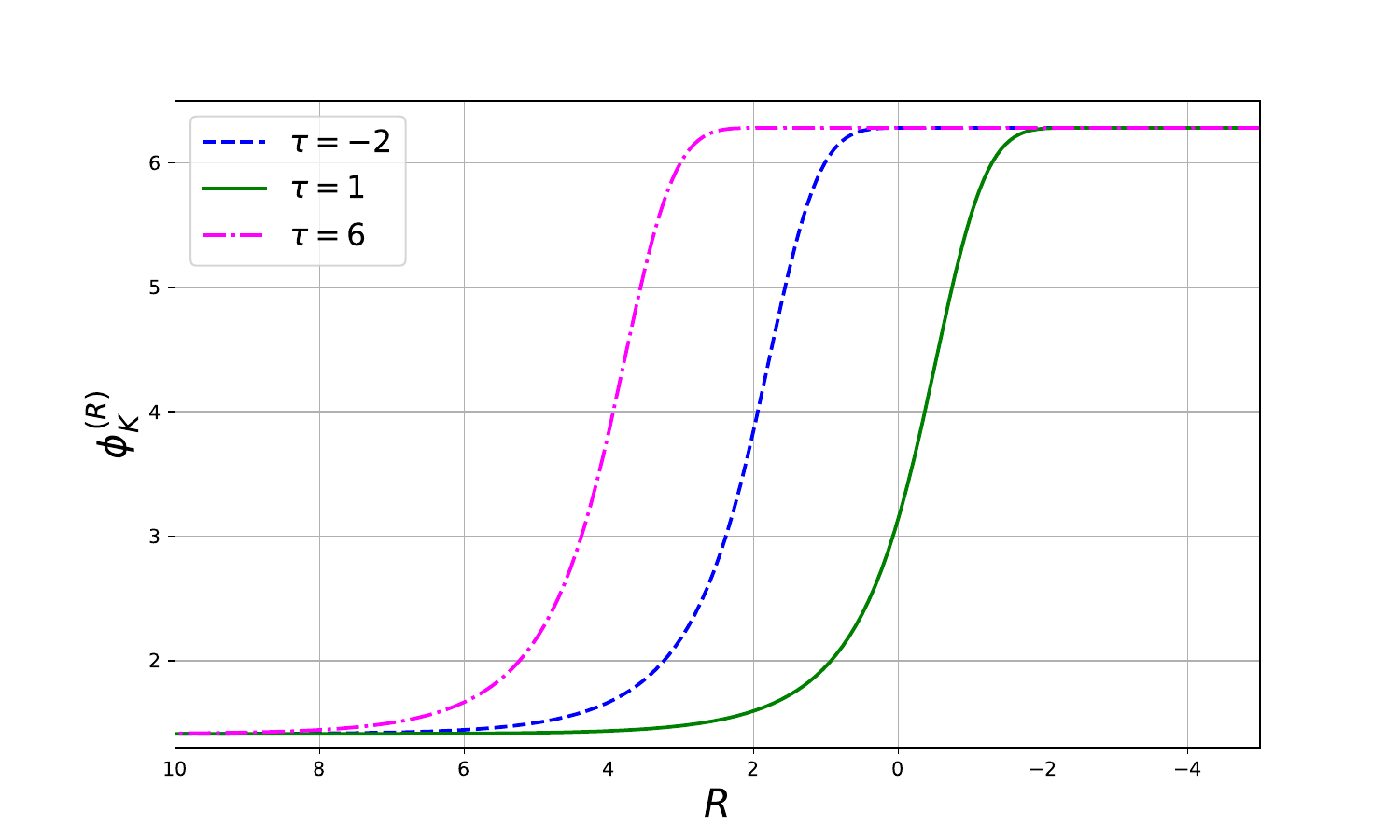}
    \caption{The kink solutions in the two wedges. Note that the $R$ coordinate is reversed in the right plot. It looks like that the kink in Minkowksi spacetime ``splits" into two kinks in Rindler wedges. Here, we set the prameters $m_0$, $\be$, $\ga_0$ and $\th_0$ to be unit.}
    \label{fig:f1}
\end{figure}

The speed of the soliton is obtained by demanding $d\ga=0$
\begin{equation}
    \frac{dR}{d\tau}=\tanh(\be\tau-\th_0).
\end{equation}
In the far past, the soliton is moving in the negative $R$ direction at the speed of light $R\sim-\tau$. It decelerates and stops at $\tau=\th_0/\be$. After that, it accelerates with time in the reverse direction, till the speed of light $R\sim\tau$.

The anti-kink solutions in the wedges are
\begin{equation}
    \phi^{(R)}_{\bar{\text{K}}}(\tau, R)=4\arctan e^{-\eta+\eta_0},
\end{equation}
\begin{equation}
    \phi^{(L)}_{\bar{\text{K}}}(\tau, R)=4\arctan e^{\eta+\eta_0},
\end{equation}
where $\eta$ is of the same form as $\ga$.

\section{The topological charge}
\label{sec:topcharge}

At a fixed time, the asymptotical values of the field are:
\begin{equation}
    \phi^{(R)}_{\text{K}}(\tau, R\rightarrow-\infty)\rightarrow2\pi, \quad \phi^{(R)}_{\text{K}}(\tau, R\rightarrow\infty)\rightarrow 2\pi\ka_{\text{K}},
\end{equation}
\begin{equation}
    \phi^{(L)}_{\text{K}}(\tau, R\rightarrow-\infty)\rightarrow0, \quad \phi^{(L)}_{\text{K}}(\tau, R\rightarrow\infty)\rightarrow 2\pi\ka_{\text{K}},
\end{equation}
and
\begin{equation}
    \phi^{(R)}_{\bar{\text{K}}}(\tau, R\rightarrow-\infty)\rightarrow0, \quad \phi^{(R)}_{\bar{\text{K}}}(\tau, R\rightarrow\infty)\rightarrow 2\pi\ka_{\bar{\text{K}}},
\end{equation}
\begin{equation}
    \phi^{(L)}_{\bar{\text{K}}}(\tau, R\rightarrow-\infty)\rightarrow2\pi, \quad \phi^{(L)}_{\bar{\text{K}}}(\tau, R\rightarrow\infty)\rightarrow 2\pi\ka_{\bar{\text{K}}},
\end{equation}
where
\begin{equation}
    \ka_{\text{K}}=\frac{2}{\pi}\arctan e^{-\ga_0}, \quad \ka_{\bar{\text{K}}}=\frac{2}{\pi}\arctan e^{\eta_0}.
\end{equation}

For the theory, we can define a topological current as
\begin{equation}
    j^\mu=\frac{1}{2\pi}\varepsilon^{\mu\nu}\pa_\nu\phi,
\end{equation}
where $\varepsilon^{\mu\nu}=(1/\sqrt{-g})[\mu\,\nu]$ is the Levi-Civita tensor. The current is conserved $\nabla_\mu j^\mu=0$. 
Then the topological charge for the soliton is given by
\begin{equation}
    Q=\int d\Si_\tau j^\tau=\frac{1}{2\pi}\int_{-\infty}^{\infty}\partial_R\phi dR,
\end{equation}
where $d\Si_\tau$ is the directed area element for the constant time slice
\begin{equation}
    d\Si_\tau=n_\tau e^{-\be R}dR.
\end{equation}
The normal vector $n_\mu=-\nabla_\mu\tau/\sqrt{|\nabla\tau\cdot\nabla\tau|}$. The charge flux is
\begin{equation}
   F_c=\int d\Si_R j^R=-\frac{1}{2\pi}\int_{-\infty}^\infty \pa_\tau\phi d\tau=0.
\end{equation}
The net charge flux is zero as the soliton is emitted and then goes back to the horizon.

With the boundary values, the topological charges of the Rindler solitons can be calculated. It is easy to find that they satisfy the relation
\begin{equation}
    Q_{\text{K}}^{(M)}=Q_{\text{K}}^{(L)}-Q_{\text{K}}^{(R)}=1,
\end{equation}
where
\begin{equation}
    Q_{\text{K}}^{(R)}=\ka_{\text{K}}-1, \quad Q_{\text{K}}^{(L)}=\ka_{\text{K}}.
\end{equation}
It looks that the solitons in Minkowski spacetime ``splits" into two solitons in Rindelr wedges, dividing the toplogical charge $Q_{\text{K}}^{(M)}$ into two. Note the negative sign before $Q_{\text{K}}^{(R)}$ arises from the fact that the $R$ direction is reversed with respect to $x$.

We have a similar relation for anti-kinks. From the solutions and charges, we can see that the kink and anti-kink solutions in the two wedges are interchangable:
\begin{equation}
    \phi^{(R)}_{\text{K}}\leftrightarrow\phi^{(L)}_{\bar{\text{K}}}, \quad \phi^{(L)}_{\text{K}}\leftrightarrow\phi^{(R)}_{\bar{\text{K}}},
\end{equation}
if we identify $\ga_0=-\eta_0$.

\section{The energy}
\label{sec:energy}

\begin{figure}[H]
    \center
    \includegraphics[width=0.95\columnwidth]{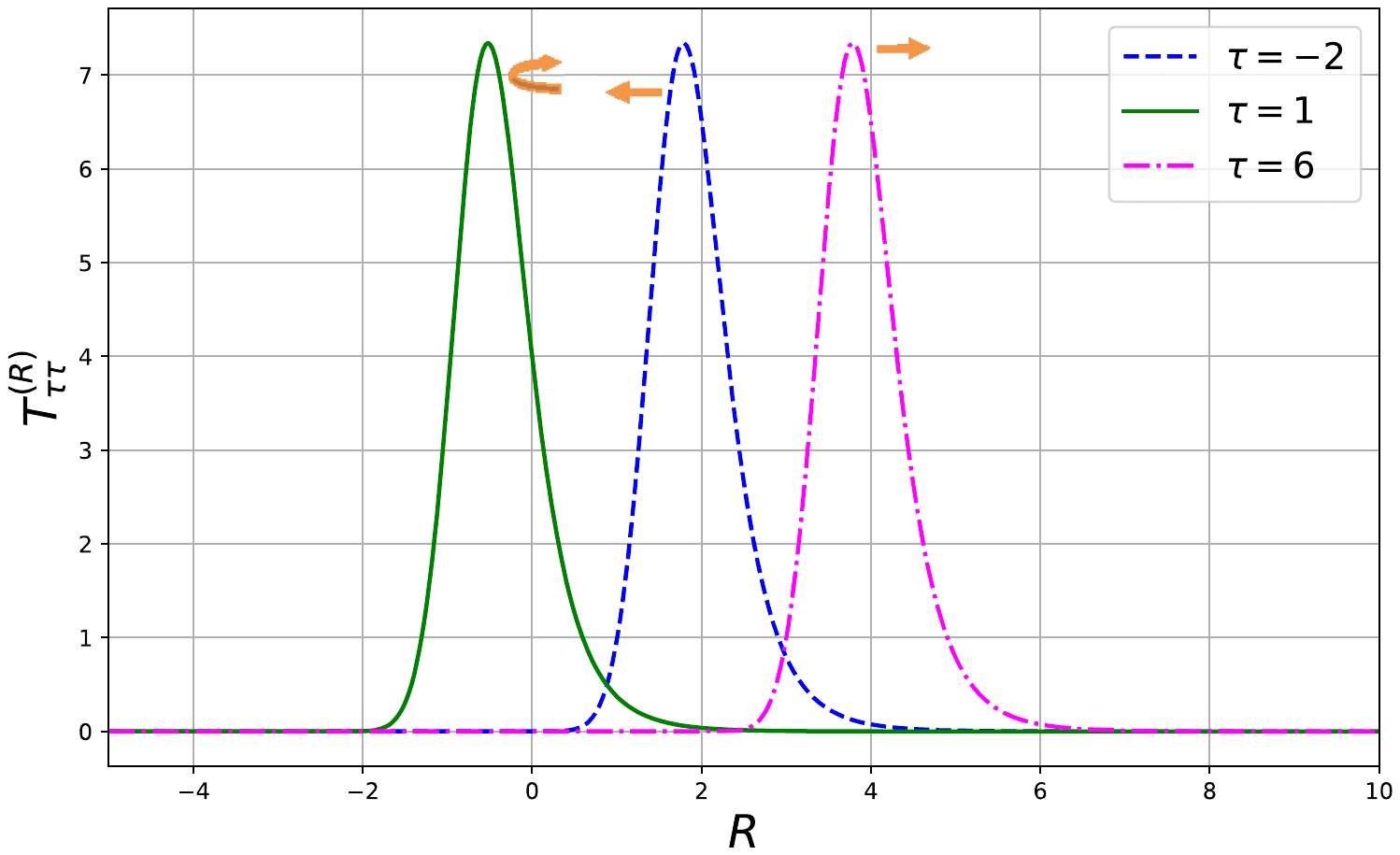}
    \caption{The distribution of $T^{(R)}_{\tau\tau}$ at different $\tau$ for the kink solution $\phi_\text{K}^{(R)}$ shown in the previous figure. The position of the peak at $\tau=1$ is the minimum coordinate $R_m$. The soliton moves from $R\rightarrow\infty$ (the horizon) and stops at $\tau=1$. Then it turns back and acceleratees towards $R\rightarrow\infty$.}
    \label{fig:f2}
\end{figure}

For the kink solution on the right wedge, the components of the energy-momentum tensor are
\begin{equation}
    T_{\tau\tau}^{(R)}=\frac{1}{2}\left[(\pa_\tau\phi)^2+(\pa_R\phi)^2\right]+e^{-2\be R}V=4\be^2\ga^2\operatorname{sech}^2(\ga-\ga_0),
\end{equation}
\begin{equation}
    T_{\tau R}^{(R)}=T_{R\tau}^{(R)}=\pa_\tau\phi\pa_R\phi=-4\be^2\tanh(\be\tau-\th_0)\ga^2\operatorname{sech}^2(\ga-\ga_0),
\end{equation}
\begin{equation}
    T_{RR}^{(R)}=\frac{1}{2}\left[(\pa_\tau\phi)^2+(\pa_R\phi)^2\right]-e^{-2\be R}V=4\be^2\tanh^2(\be\tau-\th_0)\ga^2\operatorname{sech}^2(\ga-\ga_0).
\end{equation}
The components all vanish at $\ga\rightarrow0$ and $\ga\rightarrow\infty$. The energy density has one peak within $\ga\in(0,\,\infty)$, as shown in Fig.\ (\ref{fig:f2}). The peak is located where $\ga$ satisfies
\begin{equation}
    \ga\tanh(\ga-\ga_0)=1.
\end{equation}
This euqation can be solved numerically. The peak of the energy density can vary only within the region:
\begin{equation}
    R^{\text{peak}}\geq R_m.
\end{equation}
where $R_m$ is the minimum position of the peak, which occurs when the velocity of the soliton is zero: $\tau=\th_0/\be$ so that $\ga=(m_0/\be)e^{-\be R_m}$. For $\ga_0=0$, the minimum position is at $R_m\simeq-(1/\be)[0.18+\ln(\be/m_0)]$. The soliton will turn around when the peak reaches this minimum position.

The conserved energy on a constant time slice is
\begin{equation}
    E^{(R)}=\int T_{\mu\tau}^{(R)}\xi^\mu d\Si^\tau
=\int_{-\infty}^\infty T_{\tau\tau}^{(R)}dR=4\be\ln\left(1+e^{2\ga_0}\right),
\end{equation}
where $\xi^\mu=(1,\,0)$ is the Killing vector. Unlike the Minkowski case, the energy is not essentially related to the mass $m_0$, in particular for the centered soliton $\ga_0=0$. It is also irrelevant to the velocity of the soliton and is finite at the speed of light. It is relying on $\ga_0$ and is infinite when $\ga_0\rightarrow\infty$. On the other side, when $\ga_0$ is negative, then the energy becomes negligible quickly. 

The energy flowing through a point at any $R$ in all time is given by
\begin{equation}
    F_e^{(R)}=\int T_{\mu R}^{(R)}\xi^\mu d\Si^R=-\int_{-\infty}^\infty T_{\tau R}^{(R)}d\tau=8\be\ga_0.
\end{equation}
When $\ga_0=0$, the net energy flux is zero, which means that the energy flowing in and out at any $R$ cancels out. The Rindler wedge loses energy to the horizon when $\ga_0>0$ and gains energy from the horizon when $\ga_0<0$. 

Thus, $F_e^{(R)}<E^{(R)}$ for either case. This means that an amount of energy remains as the soliton goes back the horizon $R\rightarrow\infty$. That is, the Rindler space retains some of the energy permanently. Only for the case $\ga_0\gg1$, almost all energy flows to the horizon: $E^{(R)}\simeq8\be\ga_0=F^{(R)}$.

For a kink on the left wedge, the energy density is obtained with $\ga\rightarrow-\ga$
\begin{equation}
    T_{\tau\tau}^{(L)}=4\be^2\ga^2\operatorname{sech}^2(\ga+\ga_0).
\end{equation}
The distribution of $T_{\tau\tau}^{(L)}$ is similar to the case of $T_{\tau\tau}^{(R)}$ shown in Fig.\ (\ref{fig:f2}), with a different minimum peak position $R_m$. The conserved energy and enrgy flux are respectively
\begin{equation}
    E^{(L)}=4\be\ln\left(1+e^{-2\ga_0}\right), \quad F_e^{(L)}=-8\be\ga_0.
\end{equation}
The obvious relation $F_e^{(L)}+F_e^{(R)}=0$ idicates that the energy gained in one wedge is that lost in the other wedge, i.e., energy can flow into and out the Rindler horizon. Whether energy is gained or lost depends on the relative value between the energies in the two wedges
\begin{equation}
    E^{(R)}-E^{(L)}=F_e^{(R)}=-F_e^{(L)}.
\end{equation}

If the center of the soliton deviates to the positive side, the soliton on the right wedge contains more energy and the one on the left wedge contains less. The other way occurs if the divation is negative. But the total energy is symmtric on $\ga_0$
\begin{equation}
    E^{(R)}+E^{(L)}=8\be\ln\left(2\cosh\ga_0\right).
\end{equation}
The total energy remains the same for equal $|\ga_0|$. When the center of the soliton is at the orign $\ga_0=0$, the total energy takes the minimum value: $(8\ln2)\be$, which is irrelevant to $m_0$ as well.

\section{Temporally asymptotical behaviour}
\label{sec:tempasy}

Let us consider the behaviour at lage $|\tau|$. At early time with $\tau\ll-1/\be$,
\[\ga\simeq\frac{m_0}{2\be}e^{-2\be\xi^++\th_0}=m_0e^{\th_0}v\geq 0.\]
As stated above, the soliton is emitted in the far past with the speed of light $\xi^-=\text{const}$ from the horizon. Then the kink solution takes the following form on the null coordinate in Minkowski spacetime
\[\phi^{(R)}_{\text{K}}(\tau, R)\simeq 4\arctan e^{m_0e^{\th_0}v-\ga_0}.\]
They describe one-dimensional lines along the null direction $\xi^+$ or $v$. The energy-momentum tensor is
\[T_{\mu\nu}^{(R)}\propto 4\be^2m_0^2e^{2\th_0}v^2\operatorname{sech}^2(m_0e^{\th_0}v-\ga_0).\]

At late time with $\tau\gg1/\be$, $\xi^+=\text{const}$ and
\[\ga\simeq\frac{m_0}{2\be}e^{-2\be\xi^--\th_0}=m_0e^{-\th_0}u\geq 0.\]
The kink solution is
\[\phi^{(R)}_{\text{K}}(\tau, R)\simeq 4\arctan e^{m_0e^{-\th_0}u-\ga_0}.\]
Thus, in the far future, the soliton moves toward the horizon with the speed of light, freezing on the null coordinate $\xi^-$ or $v$. The energy momentum tensor is approximately
\[T_{\mu\nu}^{(R)}\propto 4\be^2m_0^2e^{-2\th_0}u^2\operatorname{sech}^2(m_0e^{-\th_0}u-\ga_0).\]

\begin{figure}[H]
    \center
    \includegraphics[width=0.95\columnwidth]{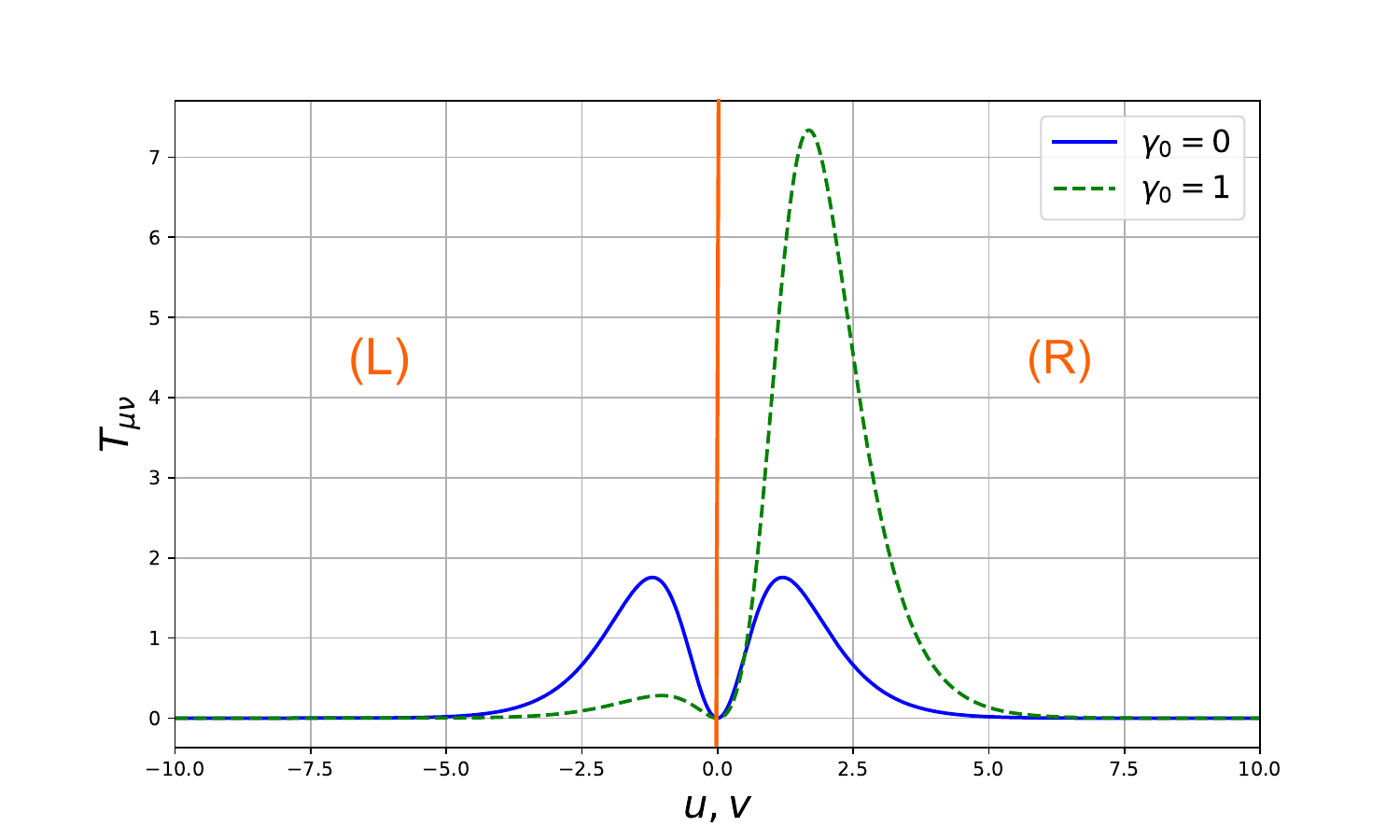}
    \caption{The distribution of energy-momentum tensor with $\th_0=0$ at large $\tau$ on null coordinates in the left (L) and right (R) wedges.}
    \label{fig:f3}
\end{figure}

With the solutions on the left side $\ga\leq0$, these form the solutons on the full null coordinates $v$ and $u$. The distribution of the energy-momentum tensor is given in Fig.\ (\ref{fig:f3}). 

\section{Discussion}
\label{sec:discussion}

The soliton in Minkowski spacetime splits into two solitons in the Rindler wedges. Each solution describes a soliton incoming from the horizon with the speed of light decelerates to stop and then turns back, accelerating to the speed of light toward the horizon. This process is analogous to an object thrown upward in the gravitational field near the ground where the acceleration is constant. It is puzzling that the soliton energy depends only on the parameter $\ga_0$ and remains finite at the speed of light, unlike in the Minkowski case. 

The soliton with larger energy gains energy from the horizon while the lower energy one loses energy. But the total energy flux of the two wedges vanishes. By constrast, the energy flux of a point particle moving in Rindler space is equal to its energy 
\begin{equation}
    F_e=E=\frac{m_0e^{-\be R}}{\sqrt{1-\dot{R}^2}}.
\end{equation}
There is no energy through the horizon. This is the same case as a soliton in Minkowski spacetime. But it is unknown why the situation for a structured object in Rindler space is quite different.

\bibliographystyle{JHEP}
\bibliography{b}

\providecommand{\href}[2]{#2}\begingroup\raggedright\begin{thebibliography}{10}

\bibitem{Manton:2004tk}
N.S.~Manton and P.~Sutcliffe, \emph{{Topological solitons}}, Cambridge
  Monographs on Mathematical Physics, Cambridge University Press (2004),
  \href{https://doi.org/10.1017/CBO9780511617034}{10.1017/CBO9780511617034}.

\bibitem{Shnir:2018yzp}
Y.M.~Shnir, \emph{{Topological and Non-Topological Solitons in Scalar Field
  Theories}}, Cambridge University Press (7, 2018).

\bibitem{Kivshar:1989ue}
Y.T.~Kivshar and B.A.~Malomed, \emph{{Dynamics of Solitons in Nearly Integrable
  Systems}}, \href{https://doi.org/10.1103/RevModPhys.61.763}{\emph{Rev. Mod.
  Phys.} {\bfseries 61} (1989) 763}.

\bibitem{Kundu:2007jh}
A.~Kundu, \emph{{Shape changing and accelerating solitons in integrable
  variable mass sine-Gordon model}},
  \href{https://doi.org/10.1103/PhysRevLett.99.154101}{\emph{Phys. Rev. Lett.}
  {\bfseries 99} (2007) 154101}
  [\href{https://arxiv.org/abs/hep-th/0703219}{{\ttfamily hep-th/0703219}}].

\bibitem{Hoare:2020fye}
B.~Hoare, N.~Levine and A.A.~Tseytlin, \emph{{Sigma models with local
  couplings: a new integrability -- RG flow connection}},
  \href{https://doi.org/10.1007/JHEP11(2020)020}{\emph{JHEP} {\bfseries 11}
  (2020) 020} [\href{https://arxiv.org/abs/2008.01112}{{\ttfamily
  2008.01112}}].

\bibitem{Aguirre:2026jbx}
A.R.~Aguirre, H.~Blas, H.F.~Callisaya and M.C.~de~Oliveira,
  \emph{{Variable-mass sine-Gordon with point defects: integrability, soliton
  transmission, and quasi-conservation}},
  \href{https://arxiv.org/abs/2608.17211}{{\ttfamily 2608.17211}}.

\bibitem{Gorbach_2007}
A.V.~Gorbach and D.V.~Skryabin, \emph{Theory of radiation trapping by the
  accelerating solitons in optical fibers},
  \href{https://doi.org/10.1103/physreva.76.053803}{\emph{Physical Review A}
  {\bfseries 76} (2007) } [\href{https://arxiv.org/abs/0707.1598}{{\ttfamily
  0707.1598}}].

\bibitem{Kundu_2008}
A.~Kundu, \emph{Exact accelerating solitons in nonholonomic deformation of the
  kdv equation with a two-fold integrable hierarchy},
  \href{https://doi.org/10.1088/1751-8113/41/49/495201}{\emph{Journal of
  Physics A: Mathematical and Theoretical} {\bfseries 41} (2008) 495201}
  [\href{https://arxiv.org/abs/0806.2743}{{\ttfamily 0806.2743}}].

\bibitem{Fac_o_2013}
M.~Facão, M.I.~Carvalho and P.~Almeida, \emph{Accelerating solitons in
  gas-filled hollow-core photonic crystal fibers},
  \href{https://doi.org/10.1103/physreva.87.063803}{\emph{Physical Review A}
  {\bfseries 87} (2013) } [\href{https://arxiv.org/abs/1304.3034}{{\ttfamily
  1304.3034}}].

\bibitem{Mukherjee_2019}
A.~Mukherjee and A.~Kundu, \emph{Novel nonlinear wave equation: Regulated rogue
  waves and accelerated soliton solutions},
  \href{https://doi.org/10.1016/j.physleta.2018.12.023}{\emph{Physics Letters
  A} {\bfseries 383} (2019) 985–990}
  [\href{https://arxiv.org/abs/1808.02293}{{\ttfamily 1808.02293}}].

\bibitem{Qin_2019}
J.~Qin, Z.~Liang, B.A.~Malomed and G.~Dong, \emph{Tail-free self-accelerating
  solitons and vortices},
  \href{https://doi.org/10.1103/physreva.99.023610}{\emph{Physical Review A}
  {\bfseries 99} (2019) } [\href{https://arxiv.org/abs/1901.02325}{{\ttfamily
  1901.02325}}].

\bibitem{Sakaguchi_2019}
H.~Sakaguchi and B.A.~Malomed, \emph{Interactions of solitons with positive and
  negative masses: Shuttle motion and coacceleration},
  \href{https://doi.org/10.1103/physreve.99.022216}{\emph{Physical Review E}
  {\bfseries 99} (2019) } [\href{https://arxiv.org/abs/1901.11243}{{\ttfamily
  1901.11243}}].

\bibitem{Melnikov:2020ret}
I.V.~Melnikov, C.~Papageorgakis and A.B.~Royston, \emph{{Accelerating
  solitons}}, \href{https://doi.org/10.1103/PhysRevD.102.125002}{\emph{Phys.
  Rev. D} {\bfseries 102} (2020) 125002}
  [\href{https://arxiv.org/abs/2007.11028}{{\ttfamily 2007.11028}}].

\bibitem{Malomed_2022}
B.A.~Malomed, \emph{Self-accelerating solitons}, {\emph{Europhysics Letters}
  {\bfseries 140} (2022) 22001}
  [\href{https://arxiv.org/abs/2210.01507}{{\ttfamily 2210.01507}}].

\bibitem{Birrell:1982ix}
N.D.~Birrell and P.C.W.~Davies, \emph{{Quantum Fields in Curved Space}},
  Cambridge Monographs on Mathematical Physics, Cambridge University Press,
  Cambridge, UK (1982),
  \href{https://doi.org/10.1017/CBO9780511622632}{10.1017/CBO9780511622632}.

\bibitem{Fulling:1989nb}
S.A.~Fulling, \emph{{Aspects of Quantum Field Theory in Curved Space-time}},
  vol.~17 (1989).

\end{thebibliography}\endgroup

\end{document}